\documentstyle[preprint,revtex,eqsecnum]{aps}
\begin{document}
\draft
\def\bra#1{\langle #1 |}
\def\ket#1{| #1 \rangle}
\def\braket#1#2{\langle #1 | #2 \rangle}
\def\annihilate{{\hat a}}
\def\create{{{\hat a}^\dagger}}
\def\shat{{\hat S}}
\def\xp{{x^\prime}}
\def\Qp{{Q^\prime}}
\def\delQ{{\delta Q}}
\def\delQp{{\delta\Qp}}
\def\delx{{\delta x}}
\def\delxp{{\delta\xp}}
\def\delX{{\delta X}}
\def\delxi{{\delta\xi}}
\def\Ss{{S_{\rm sys}}}
\def\Sr{{S_{\rm res}}}
\def\e{{\rm e}}
\def\re{{\rm Re}}
\def\im{{\rm Im}}
\def\intof{\int_{t_0}^{t_f}}
\def\intot{\int_{t_0}^t}
\def\rhot{{\tilde \rho}}
\def\dLdX{{{\partial L}\over{\partial X}}}
\def\dLdXd{{{\partial L}\over{\partial {\dot X}}}}
\def\Pinv{P_{\rm inv}}
\def\winv{w_{\rm inv}}

\begin{title}
QUANTUM DISSIPATIVE CHAOS
\end{title}

\author{{\bf Todd A. Brun}}
\begin{instit}
Department of Physics \\
California Institute of Technology \\
Pasadena, CA  91125
\end{instit}

\nonum\section{Abstract}
\begin{abstract}
Using the decoherence formalism of Gell-Mann and Hartle, a quantum
system is found which is the equivalent of the classical dissipative
chaotic Duffing oscillator.  The similarities and
differences from the classical oscillator are examined; in particular,
a new concept of quantum maps is introduced, and alterations in the classical
strange attractor due to the presence of scale-dependent quantum
effects are studied.  Classical quantities such as the Lyapunov exponents
and fractal dimension are examined, and quantum analogs are suggested.
These results are generalized into a framework for quantum dissipative chaos,
and there is a a brief discussion of other work in this area.
\end{abstract}

\vfil\eject

\section{{\bf INTRODUCTION}}

Since classical chaos first began to be studied, a conspicuous puzzle
has been how to reconcile this nonlinear, purely classical phenomenon
with an underlying linear quantum theory.  If we believe, as we must to be
consistent, that all of physics is fundamentally quantum mechanical in
nature, then we must further believe that true chaotic systems do not
exist.  At some point, at length scales determined by Planck's
constant, the deterministic uncertainties of classical chaos
must give way to the
probabilistic uncertainties of quantum mechanics.

But tackling these problems is not simple.  The nonlinear equations of
chaos are, in general, only solvable with modern high-speed computers,
and their quantum analogs share this limitation.  Also, chaos itself
encompasses two major types of behavior:  Hamiltonian chaos in which
energy is conserved, and dissipative chaos.  It
is in principle straightforward to find
and solve quantum equivalents to Hamiltonian
systems, if difficult in practice, and considerable progress has
been made in recent years in understanding these systems.
Dissipative systems are
much more foreign to quantum mechanics as it is usually studied.

Recently, Murray Gell-Mann and James Hartle have used the decoherence
functional formalism of quantum mechanics to show how quasiclassical
laws can arise from an underlying quantum theory
\cite{GMHart1,GMHart2,GMHart3}.
I applied this approach
to the problem of Brownian motion, demonstrating how their scheme
reproduces exactly the classical Langevin equation in a fairly broad class
of systems \cite{Brun}.

A natural next step is to apply this to systems with interesting
classical behavior.
Since dissipation is easily and indeed naturally
included in such systems, an obvious candidate for study is dissipative chaos.
Once a quantum system is found whose limiting behavior is equivalent to
a classical chaotic system, we can study how
the residual quantum mechanical effects alter
the system, and what difference this makes to the classical behavior.

In section II we briefly examine the family of quantum systems from which
we will draw our model, and derive the quasiclassical equations of motion
for them.  We then go to the limit of an infinite reservoir of oscillators
with a continuum of frequencies, and specialize to the case of a forced,
damped nonlinear oscillator.

In section III we examine the classical behavior of one such system,
the damped, driven Duffing oscillator.  There is a brief discussion of
dissipative chaos, the structure of the strange attractor, and the
bifurcations leading to chaos.  Several quantities useful
for characterizing the chaotic behavior are defined:  the fractal dimension
and Lyapunov exponents, and their relationships are examined.

In section IV we look at the decoherence functional, and define the idea of
a quantum map.  The system is examined as a Wigner distribution, and we
see how the invariant measure of the strange attractor goes over to the
quantum case.  Problems of coarse-graining and decoherence are discussed.
Then we look at the system from a Master equation point of
view, and compare this description to the decoherence functional approach.
In section V we see how the various classical quantities
used to characterize chaotic behavior can be reinterpreted for our quantum
system, by treating it as a classical system with noise for sufficiently
coarse length scales.

A few other treatments of quantum dissipative chaos are
mentioned in section VI, and the differences between Hamiltonian and
dissipative chaos are pointed out.  Finally, in section VII a case is
made for a general theory of quantum dissipative chaos.

\section{{\bf DAMPED DRIVEN QUANTUM SYSTEMS.}}

\subsection{The Quantum Systems.}

Picking a good set of candidate systems requires some thought.
Many widely studied sets of chaotic equations have only a loose connection
to actual physical systems; many others are extreme coarse grainings of
very complicated systems with many degrees of freedom, e.g., fluid
dynamics.  It is much better to deal with comparatively simple systems,
whose decoherence functionals can be calculated easily.  For this reason,
I have elected to study damped driven nonlinear oscillators, which can
be easily modelled as particles moving in a potential well, interacting
with a reservoir of simple linear oscillators.  In particular, I will
concentrate on one such system, the damped, driven Duffing oscillator.

Earlier work has chiefly considered systems with reservoirs
in a thermal state.  For the purpose of this model, I wish to consider
instead a system whose reservoir is initially in a coherent state.

Consider a system of $N$ harmonic oscillators.  We assume them to be
in a state $\ket{\{\nu\}}$, where $\{\nu\}$ represents a set of $N$
complex numbers $\nu_j$.
If $\annihilate_i$ is the annihilation operator
for the $i$th oscillator, then $\annihilate_i \ket{\{\nu\}} = \nu_i
\ket{\{\nu\}}$.

As shown in earlier papers \cite{GMHart3,Brun},
the decoherence functional for a system
interacting with a reservoir is
\begin{eqnarray}
D[\xp(t),x(t)] =&& \exp\biggl\{ i(\Ss[\xp(t)] - \Ss[x(t)])
  /\hbar \biggr\}
  \int \delQp \delQ\ \delta(\Qp(t_f) - Q(t_f)) \nonumber \\
&&\ \ \times \exp\biggl\{ i (\Sr[\Qp(t)] - \Sr[Q(t)] \nonumber \\
&&\ \ \ - \int_{t_0}^{t_f} (V(\xp(t),\Qp(t)) - V(x(t),Q(t))) dt)/\hbar
  \biggr\}  \nonumber \\
&&\ \ \times \rho(\xp_0,\Qp_0; x_0,Q_0) \nonumber \\
=&&\exp\biggl\{ i(\Ss[\xp(t)] - \Ss[x(t)] + W[\xp(t),x(t)])/\hbar\biggr\}
  \rhot(\xp_0; x_0).
\end{eqnarray}
Here $\Ss[x(t)]$ is the action of the system for a given
trajectory $x(t)$, $\Sr[Q(t)]$ is the action of the reservoir
for a given trajectory $Q(t)$, and $V(x,Q)$ is the interaction
potential between the system and reservoir variables.  We will assume
this to be a bilinear potential of the form
\begin{equation}
V(x,Q) = - \sum_k \gamma_k x Q^k,
\label{interaction}
\end{equation}
where $Q^k$ is the coordinate of the $k$th oscillator in the resevoir.
We will shortly allow the number of oscillators to go to infinity,
and assume a continuum of oscillator frequencies, but for now
let us deal with the discrete case.

We will also assume that the density matrix factors:
\begin{equation}
\rho(\xp_0,\Qp_0; x_0,Q_0) = \chi(\xp_0; x_0)\phi(\Qp_0; Q_0),
\end{equation}
where $\phi(\Qp_0; Q_0) = \braket{\Qp_0}{\{\nu\}}\braket{\{\nu\}}{Q_0}$
is the pure coherent state described above.

We can readily calculate the influence functional for this system.
It is just
\begin{eqnarray}
\exp\biggl\{ i W[\xp(t),x(t)]/\hbar \biggr\} = &&
  \int\int\int dQ_f dQ_0 d\Qp_0 K^*_{\xp(t)}(Q_f; \Qp_0) K_{x(t)}(Q_f; Q_0)
  \phi(\Qp_0; Q_0) \nonumber \\
= && \bra{\{\nu\}} \shat_{\xp(t)}^\dagger \shat_{x(t)} \ket{\{\nu\}},
\end{eqnarray}
where $K_{x(t)}(Q_f; Q_0)$ is the transition amplitude from $Q_0$ to
$Q_f$ of the reservoir and
$\shat_{x(t)}$ is the time evolution operator of a forced harmonic
oscillator driven by the time-dependant interaction $V(x(t),Q)$ given
in (\ref{interaction}).  This is a well-known problem \cite{Merzbacher}.
For a single oscillator of frequency
$\omega$ the operator is
\begin{equation}
\shat_{x(t)} = \exp\biggl[\alpha\create - \alpha^*\annihilate \biggr]
  = D(\alpha),
\end{equation}
where
\begin{equation}
\alpha = {{i\gamma}\over\sqrt{2m\omega\hbar}} \intof
  \e^{i\omega s} x(s) ds,
\end{equation}
and
\begin{equation}
D(\alpha) \ket{\nu} = \ket{\nu + \alpha} \e^{i\im(\alpha\nu^*)}.
\end{equation}
For $\nu = 0$ this just reduces to the usual form of the influence functional
for an oscillator initially in the ground state:
\[
\exp \biggl\{i W[\xp(t),x(t)]/\hbar \biggr\} = \exp {i\over\hbar} \biggl\{
  + {{i\gamma^2}\over{4m\omega}} \intof dt \intof ds\ \cos(\omega(t-s))
  (\xp(t) - x(t)) (\xp(s) - x(s))
\]
\begin{equation}
- {{\gamma^2}\over{2m\omega}} \intof dt \intot ds\ \sin(\omega(t-s))
  (\xp(t) - x(t)) (\xp(s) + x(s)) \biggr\}.
\end{equation}
For non-zero $\nu$ we get an additional exponent of the form
\begin{equation}
\im (2\nu\alpha^* + 2\nu^*\alpha^\prime) =
  \gamma \sqrt{{2\over{m\omega\hbar}}}
  \intof dt (\re\nu\cos(\omega t) + \im\nu\sin(\omega t)) (x(t) - \xp(t)).
\end{equation}
Generalizing this to many oscillators, we get the influence phase
\[
W[\xp(t),x(t)] = \sum_k
  {{i\gamma_k^2}\over{4m\omega_k}} \intof dt
  \intof ds \cos(\omega_k(t-s))
  (\xp(t) - x(t)) (\xp(s) - x(s))
\]
\[
- {{\gamma_k^2}\over{2m\omega_k}} \intof dt
  \intot ds \sin(\omega_k(t-s))
  (\xp(t) - x(t)) (\xp(s) + x(s))
\]
\begin{equation}
+ \gamma_k \sqrt{{{2\hbar}\over{m\omega_k}}}
  \intof dt (\re\nu_k\cos(\omega_k t) + \im\nu_k\sin(\omega_k t))
  (\xp(t) - x(t)).
\label{discrete}
\end{equation}
For practical purposes, we generally assume that the interaction began
at $t_0 = 0$ and continued up until some final time
$t_f$, so as to avoid having infinite limits in the integrals.

We will now assume that the action of the system variables is of the
usual form
\begin{equation}
\Ss[x(t)] = \intof L(x(t),{\dot x}(t)) dt.
\end{equation}
We can then change variables to
\begin{mathletters}
\begin{eqnarray}
X = && {1\over2}(x + \xp), \\
\xi = && x - \xp,
\end{eqnarray}
\end{mathletters}
and write the decoherence functional in terms of the new variables.
This is easily shown to be

\vfil\eject

\begin{eqnarray}
D[X(t),\xi(t)] = && \exp {i\over\hbar} \biggl\{ \sum_k \intof dt\
  \xi(t) \biggl[ - {d\over{dt}}\biggl(\dLdXd\biggr)(X(t),{\dot X}(t)) +
  \dLdX(X(t),{\dot X}(t)) \nonumber \\
&& + \gamma_k\sqrt{{2\hbar}\over{m\omega_k}}
  \bigl(\re\nu_k \cos(\omega_k(t)) + \im\nu_k \sin(\omega_k(t))\bigr)
  - {{\gamma_k^2}\over{m\omega_k}} \intot ds
  \sin(\omega_k(t-s)) X(s) \biggr] \nonumber \\
&& + {{i\gamma_k^2}\over{4m\omega_k}} \intof dt \intot ds
  \cos(\omega_k(t-s)) \xi(t)\xi(s) \nonumber \\
&&  - \xi_0 \dLdXd(X_0,{\dot X}_0) + O(\xi^3) \biggr\}
  \chi(\xp_0; x_0).
\end{eqnarray}
Note that the real part of the phase includes the Euler-Lagrange
equation of motion for the system, with the addition of a retarded
force due to the interaction with the reservoir.  The imaginary part
is strictly non-negative, with a minimum at $\xi(t) \equiv 0$, and hence
tends to suppress $D[\xp(t),x(t)]$ for large $\xi$.  This makes
our expansion in terms of $\xi(t)$ seem reasonable, and also causes
the system to decohere, at least approximately, since $\xi(t) \ne 0$
corresponds to off-diagonal terms.  The $\xi_0$ term occurs because of
an integration by parts.

\subsection{The Classical Equivalent.}

We can now look at the classical system equivalent to the above quantum
system, i.e., with the same action functional and distribution of
oscillators.  A coherent state is often characterized as a more
``classical'' state of an oscillator than the usual Fock states; it can
be thought of as the state of an oscillator begun at a given initial
position and momentum, within the limits imposed by the uncertainty
principle.

We will begin by assuming knowledge of the trajectory of the system
variable $x(t)$, and ask what the behavior of the reservoir of
harmonic oscillators will be \cite{Zwanzig}.
Assume that we start the oscillators
in a definite state $Q^k(t=t_0) = q^k, {\dot Q^k}(t=t_0) = v^k.$  The
interaction potential is linear, so we can treat the trajectory of the
system variable $x(t)$ as a simple driving force, giving us an equation
of motion for the $k$th oscillator
\begin{equation}
{{d^2 Q^k}\over{dt^2}} = - \omega_k^2 Q^k + (\gamma_k/m) x(t).
\end{equation}
The solution to this equation is simply
\begin{eqnarray}
Q^k(t) = && q^k \cos(\omega_k(t-t_0))
  + (v^k/\omega_k) \sin(\omega_k(t-t_0)) \nonumber\\
&& + {{\gamma_k}\over{m\omega_k}} \intot \sin(\omega_k(t-s)) x(s) ds.
\label{Qkexpression}
\end{eqnarray}

If the system is described by a Lagrangian $L(x,{\dot x})$, then we can
write down the Euler-Lagrange equation
\begin{equation}
{d\over{dt}}\dLdXd(x(t),{\dot x}(t)) - \dLdX(x(t),{\dot x}(t))
  + \sum_k \gamma_k Q^k(t) = 0.
\end{equation}
We can clearly substitute the above expression (\ref{Qkexpression}) for
$Q^k(t)$ in the Euler-Lagrange equation to get
\[
{d\over{dt}}\dLdXd(x(t),{\dot x}(t)) - \dLdX(x(t),{\dot x}(t))
  + \sum_k \gamma_k \biggl(
  q^k \cos(\omega_k(t-t_0))
  + (v^k/\omega_k) \sin(\omega_k(t-t_0))
\]
\begin{equation}
  + {{\gamma_k}\over{m\omega_k}} \intot \sin(\omega_k(t-s)) x(s) ds \biggr) =
0.
\end{equation}
This expression is clearly closely related to the real part of the
phase in the decoherence functional, if we make the identity
\begin{mathletters}
\begin{eqnarray}
\sqrt{{2\hbar}\over{m\omega_k}} \re\nu_k = && q^k, \\
\sqrt{{2\hbar\omega_k}\over m} \im\nu_k = && v^k.
\end{eqnarray}
\end{mathletters}
If we write the above classical equation as $e(t) = 0$, then the
real part of the phase is just
\[
\intof \xi(t) e(t) dt.
\]

How do we interpret the imaginary part of the phase, however?  In
treating reservoirs in an initial thermal state, we identified this
term as the effect of a stochastic force $F(t)$ arising due to
thermal noise.  However, in this system, there {\it is} no noise
classically.  The persistence of this term indicates a fundamental difference
between the quantum and classical systems.  As Gell-Mann and Hartle
point out \cite{GMHart3}, in the quantum system
there is always noise from zero-point oscillations,
unlike classical oscillators.  So the actual equation of motion derived
from the quantum theory is
\[
0 = {d\over{dt}}\dLdXd(x(t),{\dot x}(t)) - \dLdX(x(t),{\dot x}(t))
  - F(t)
\]
\begin{equation}
+ \sum_k \gamma_k \biggl(
  q^k \cos(\omega_k(t-t_0))
  + (v^k/\omega_k) \sin(\omega_k(t-t_0))
  - {{\gamma_k}\over{m\omega_k}} \intot \sin(\omega_k(t-s)) x(s) ds \biggr).
\end{equation}
where $F(t)$ is a stochastic force with $\langle F(t) \rangle = 0$, and
a two-time correlation function
\begin{equation}
\langle F(t)F(s) \rangle = \sum_k {{\gamma_k^2 \hbar}\over{4m\omega_k}}
  \cos(\omega_k(t-s)).
\end{equation}

\subsection{The Continuum Limit.}

In order to consider the sorts of classical systems we are concerned
with, we must go to the limit of a continuum of oscillator frequencies,
both classically and quantum mechanically.  In doing this, we replace
our sums over oscillators with integrals over a distribution function
$g(\omega)$.  The usual choice for such a $g(\omega)$ is the Debye
distribution\cite{Zwanzig,CaldLegg}:
\begin{equation}
g(\omega) = \cases{ \eta \omega^2/\Omega^2, & $\omega < \Omega$, \cr
  0, & $\omega > \Omega$. \cr}
\label{Debye}
\end{equation}
The reservoir degrees of freedom will become a continuum,
$Q^k(t) \rightarrow Q(\omega,t)$, and the
eigenvalues $\nu_k$ will become a
continuous complex function $\nu(\omega)$.
In general, $\Omega$ must be taken to be fairly large.  More precisely, we
want $\Omega \gg 1/(t_f - t_0)$, so that the relaxation time of the
reservoir is much less than the time-scale of the problem.

Let's consider now the various components of $W[X(t),\xi(t)]$ one at a time.
In the continuum limit, we have
\begin{equation}
\sum_k {{\gamma_k^2}\over{m\omega_k}} \intot ds \sin(\omega_k(t-s)) X(s)
  \rightarrow  {1\over{m}} \int_0^\Omega d\omega \intot ds
  {{g(\omega)}\over\omega} \sin(\omega(t-s)) X(s).
\end{equation}
We can invert the order of integration and do the $\omega$ integral,
substituting (\ref{Debye}) for $g(\omega)$:
\begin{eqnarray}
{1\over{m}} \intot ds \int_0^\Omega d\omega
  {{g(\omega)}\over\omega} \sin(\omega(t-s)) X(s)
= && {\eta\over{m\Omega^2}} \intot ds \int_0^\Omega d\omega
  \omega \sin(\omega(t-s)) X(s), \nonumber\\
= && {\eta\over{m\Omega^2}} \intot ds \biggl(
  - {{\Omega\cos(\Omega(t-s))}\over{t-s}}
  + {{\sin(\Omega(t-s))}\over{(t-s)}^2} \biggr) X(s), \nonumber\\
= && {\eta\over{m\Omega^2}} \intot ds {d\over{ds}}
  \biggl( {{sin(\Omega(t-s))}\over{t-s}} \biggr) X(s).
\end{eqnarray}

Now we use the fact that $\Omega$ is large.  This implies that the
sinusoidal terms will oscillate very rapidly, so that the integral will
tend to cancel out to zero.  We expect the largest contribution
to come in the region where $s$ is close to $t$.  Thus, we expand $X(s)$
about $t$ to get $X(s) \approx X(t) - {\dot X}(t)(t-s) + \cdots$.
Substituting this into the above integral, we can solve it term by
term to get
\begin{equation}
{1\over{m}} \intot ds \int_0^\Omega d\omega
  {{g(\omega)}\over\omega} \sin(\omega(t-s)) X(s)
  \approx - {\eta\over{m\Omega}} X(t)
  - {{\pi\eta}\over{2m\Omega^2}} {\dot X}(t) + O(\Omega^{-3}),
\label{Dissipation}
\end{equation}
where the additional terms become small in the limit of a large
cutoff $\Omega$.

The second term has the form of a dissipation with
constant $2\Gamma = \pi\eta/2mM\Omega^2$.  The first term is a linear restoring
force.  If the system were a harmonic oscillator, this would cause a
shift in the oscillator frequency.  We can absorb this term into the
system action as an additional harmonic oscillator potential:
\begin{equation}
\Ss[x(t)] \rightarrow \Ss^\prime[x(t)] = \Ss[x(t)]
  - \intof {\eta\over{2m\Omega}} x^2(t).
\label{FreqShift}
\end{equation}
If our Lagrangian is the usual $L(x,{\dot x}) = {1\over2} M {\dot x}^2
- U(x)$, then we effectively have a new Lagrangian
\begin{equation}
L(x,{\dot x}) \rightarrow L^\prime(x,{\dot x}) =
  {1\over2} M {\dot x}^2 - U^\prime(x),
\end{equation}
where
\begin{equation}
U^\prime(x) = U(x) + {\eta\over{2m\Omega}} x^2.
\end{equation}
In subsequent analysis, it will be $U^\prime(x)$ that we are
interested in, as the effective potential.

The imaginary part of $W[X,\xi]$ is also of interest.  Here we have
\[
\sum_k {{\gamma_k^2}\over{4m\omega_k}} \intof ds
\cos(\omega_k(t-s))\xi(t)\xi(s)
 \rightarrow {1\over{4m}}\int_0^\Omega d\omega \intof ds
  {{g(\omega)}\over{\omega}} \sin(\omega(t-s)) \xi(t)\xi(s),
\]
\[
= {\eta\over{4m\Omega^2}} \intof ds \int_0^\Omega d\omega
  \ \omega \cos(\omega(t-s)) \xi(t)\xi(s),
\]
\begin{equation}
= {\eta\over{4m\Omega^2}} \intof ds {d\over{ds}} \biggl(
  {{\cos(\Omega(t-s)) - 1}\over{t-s}} \biggr) \xi(t)\xi(s).
\label{Imag}
\end{equation}
If we do this derivative we see
\[
{d\over{ds}}\biggl( {{\cos(\Omega(t-s)) - 1}\over{t-s}} \biggr) =
  {{\Omega\sin(\Omega(t-s))}\over{t-s}} +
  {{\cos(\Omega(t-s)) - 1}\over{(t-s)^2}},
\]
where the first term is larger than the second by a factor of roughly
$\Omega(t_f - t_0)$.  In the limit of large $\Omega$ this term approaches
a delta function.  We again expand $\xi(s)$,
\[
\xi(s) \approx \xi(t) - {\dot \xi}(t) (t-s) + \cdots
\]
and substitute it into (\ref{Imag}) to get
\begin{equation}
{\eta\over{4m\Omega^2}} \intof ds {d\over{ds}} \biggl(
  {{\cos(\Omega(t-s)) - 1}\over{t-s}}\biggr) \xi(t)\xi(s)
\approx {{\pi\eta}\over{4m\Omega}} \xi^2(t) + O(\Omega^{-2}).
\end{equation}
We define a new constant $K = \pi\eta/2m\Omega$.

At this point, someone will likely cry foul.  In the real part, we kept
the two lowest order terms, while in the imaginary part we kept only
one!  I will give three different arguments why this is legitimate:
\smallskip

1.  The real and imaginary parts of $W[X(t),\xi(t)]$ serve different
purposes in the decoherence functional, and have different effects.
A small correction to the imaginary part has only a minor effect on the
level of decoherence.  The inclusion of a small amount of dissipation in
the real part, by contrast, leads to qualitatively different solutions.
\smallskip

2.  The leading order term of the real part can be absorbed into the
system action by going to an effective potential $U^\prime(x)$,
as we've seen, so that we must go to the next order to observe fundamental
changes in the classical equation of motion.
\smallskip

3.  The imaginary part of $W[X(t),\xi(t)]$ corresponds
to a small stochastic noise.  In the above limit the two-time correlation
function becomes $\langle F(t)F(s) \rangle =
(\pi\eta\hbar/4m\Omega) \delta(t-s) = \hbar K \delta(t-s)$.
In the usual limit where $\hbar$
is small, any corrections to the above correlation function would be
too tiny to matter.  The dissipative term lacks this factor of $\hbar$.
\smallskip

Finally, we have the terms arising from the initial condition
of the reservoir variables.  When we go to the continuum limit here,
the discrete sum in (\ref{discrete}) becomes an integral.  If we choose
$\nu(\omega) \sim \delta(\omega - \omega_0)$,
then we get
\begin{equation}
\sum_k \gamma_k\sqrt{{2\hbar}\over{m\omega_k}} (a_k\cos(\omega_k t)
  + b_k\sin(\omega_k t))\xi(t)
  \ \ \rightarrow M q \cos(\omega_0 t + \phi_0) \xi(t),
\end{equation}
where $q$ can be set arbitrarily by adjusting the amplitude of
$\re\nu(\omega)$.  This term has the form of a
periodic driving force.  We can set the phase $\phi_0$ to zero by making
$\im\nu(\omega) = 0$.
Physically, the presence of this term is equivalent
to the system being driven by a plane wave at the
frequency $\omega_0$.

Thus, our decoherence functional becomes
\begin{eqnarray}
D[X(t),\xi(t)] = && \exp {i\over\hbar} \biggl\{ \intof dt \biggl(
  - M{\ddot X}(t) - {{d U^\prime}\over{dX}}(X(t)) - 2M\Gamma {\dot X}(t)
  + M q \cos(\omega_0 t) \biggr) \xi(t) \nonumber\\
&& - M{\dot X}_0\xi_0 + i K \intof dt\ \xi^2(t)
  + O(\xi^3) \biggr\}\chi(\xp_0; x_0).
\label{cont_decoherence}
\end{eqnarray}
which gives us a quasiclassical equation of motion
\begin{equation}
{\ddot x} + {1\over M}{{d U^\prime}\over{dx}}(x) + 2\Gamma{\dot x}
  = q \cos(\omega_0 t) + F(t)/M,
\label{EOM}
\end{equation}
where $F(t)$ is a stochastic force with $\langle F(t) \rangle = 0$ and
$\langle F(t)F(s) \rangle = \hbar K \delta(t-s)$.  In a completely
classical derivation, of course, this stochastic force would be absent.
Thus, we have found a quantum system equivalent (in the appropriate limits)
to a classical nonlinear oscillator with a periodic driving force and
dissipation.  All that remains now is to specialize to a chaotic example.

\section{{\bf THE DAMPED, DRIVEN DUFFING OSCILLATOR}}

The quasiclassical equation of motion (\ref{EOM}) is a fairly general
expression for a one-dimensional damped, driven system.  Many such
systems exist which exhibit chaotic behavior for some values of the
constants $\Gamma$ and $q$.  The ordinary pendulum is an example of
such a system, where $U^\prime(x) = - \cos(x)$.
We will be examining another system:  the damped,
driven Duffing oscillator.  This nonlinear oscillator has a polynomial
potential
\begin{equation}
{1\over M}U^\prime(x) = {1\over4} x^4 - {1\over2}x^2.
\label{Potential}
\end{equation}
We choose units to set $M = 1$.
This system
has the advantage of having been thoroughly studied and examined in the
past, and also, since the potential is a polynomial, of not having an
infinite number of nonzero deriviatives.  We will later see that this is
convenient, though not vital.

The equation of motion is now
\begin{equation}
{\ddot x} + 2\Gamma {\dot x} + (x^3 - x) = q \cos(\omega_0 t) + F(t).
\end{equation}
The potential is double-welled (see fig.~1).  For some values of the constants,
the oscillator undergoes periodic motion.  By adjusting the frequency,
one causes the system to undergo a series of bifurcations
until eventually it enters into a region of chaotic behavior, typified by
the presence of a strange attractor (see fig.~2).  If one adjusts the driving
force further and further the chaotic region is eventually left, and periodic
motion returns \cite{GuckHolmes}.

It is convenient to look at the long-term chaotic behavior in terms of
a {\it constant phase} map or {\it surface of section}.
That is, we consider the position $x$ and
momentum $p$ at the discrete times $t_i = 2\pi i/\omega_0$.  By then
plotting the values $x_i$ and $p_i$ we make the fractal structure of the
strange attractor very clear (see fig.~3).  We are also able to bring the
mathematical toolbox of discrete dynamical system theory to bear on the
problem.  Formally, we define the constant phase map $x_i \rightarrow
x_{i+1} = f_x(x_i,p_i)$, $p_i \rightarrow p_{i+1} = f_p(x_i,p_i)$,
where {\bf f} is an operator which evolves the point $(x_i,p_i)$ in
phase space forward in time by $2\pi/\omega_0$.

We can now define a probability measure $P(x,p)$ on our phase space.  In
this discrete dynamics, it evolves according to the equation
\begin{equation}
P_{i+1}(x,p) = \int dx^\prime \int dp^\prime \delta(x-f_x(x^\prime,p^\prime))
  \delta(p-f_p(x^\prime,p^\prime)) P_i(x^\prime,p^\prime),
\label{Measure}
\end{equation}
It is very useful then to consider an {\it invariant
measure}, which gives a probability distribution on the strange attractor.
This is defined by the equation
\begin{equation}
P_{i+1}(x,p) = P_i(x,p) = \Pinv(x,p).
\end{equation}
For a chaotic system such as the Duffing oscillator, $\Pinv$ will not be
an analytic function; rather, it will be a generalized function.  Also,
it will not in general be unique; there are many invariant measures,
most corresponding to unstable solutions, fixed points or periodic points.
It has been shown that the inclusion of a small amount of noise removes
both of these objections, eliminating the unstable solutions and making the
function analytic \cite{Wolfetal}.  The inclusion of noise effectively
broadens the $\delta$-functions in (\ref{Measure}), making it impossible
for probability measure to remain poised on an unstable
fixed-point solution or periodic
orbit, leaving only the strange attractor as a stable set.
One can therefore define the classical $\Pinv$ as
the limit of this unique $\Pinv$ as the noise goes to zero; it will still
be a generalized function with a fractal structure, but now unique.  Note
that in the quantum system, the noise is {\it always} non-zero; we will see how
this modifies our definitions in sections IV and V.

The structure of the strange attractor arises as a limit of repeated
stretching and folding of phase space.  As we look closer and closer
at the component points of the attractor, we see repeated layers of
substructure at every scale (see fig.~4).
Such infinite substructure is commonly
characterized by its {\it fractal dimension}.  There are a number of
ways of defining dimension, each of which has slightly different
properties.

One common definition is that of the {\it capacity} or {\it Kolmogorov
dimension}, $D_C$.  This is calculated by means of a box-counting
algorithm.  Phase space is divided into small cells of linear size
$\epsilon$, and one counts the number of cells $N(\epsilon)$ which
contain points of the attractor.  The dimension is then
\begin{equation}
D_C = \lim_{\epsilon \rightarrow 0} - {{\ln N(\epsilon)}\over{\ln\epsilon}}.
\end{equation}

Though this definition is fairly easy to calculate numerically, it does not
reflect the fact that an orbit may visit regions of the attractor with
varying frequency.  To take this into account, one may instead use the
{\it information dimension}, $D_I$.  Again, phase space is divided into
cells of linear dimension $\epsilon$.  The probability that a given point
will fall in the $i$th cell is $p_i$.  $D_I$ is then
\begin{equation}
D_I = \lim_{\epsilon\rightarrow0} {{\sum\limits_i p_i\ln p_i}
  \over{\ln\epsilon}}.
\label{infodim}
\end{equation}
If $p_i$ is equal for all cells that are visited, then $D_I = D_C$; otherwise
$D_I < D_C$.

If one examines a small cell of phase space evolve according to our
equation, it will tend to be stretched along one dimension.  The overall
phase space volume, however, will contract, due to the effects of dissipation.
This stretching is what provides the well known signature of chaotic
systems, sensitivity to initial conditions.  The contraction, together
with the more global process of folding, is what leads to the fractal structure
of the attractor.  One can average these two effects over the entire
attractor to calculate the Lyapunov exponents.  In our two-dimensional
phase space, this will be a pair of numbers $\lambda_1$ and $\lambda_2$,
with $\lambda_1 > 0$ characterizing the stretching and $\lambda_2 < 0$
characterizing the contraction.  Since overall phase space volume is
shrunk by this system, clearly $\lambda_1 + \lambda_2 < 0$.  If our phase
space was $n$ dimensional, there would clearly be $n$ characteristic
exponents.

Lyapunov exponents are calculated by considering the time evolution of an
infinitesimal frame of basis vectors in phase space.  One can
perform a Gram-Schmidt orthogonalization, separating out the most rapidly
increasing direction from less rapidly increasing directions repeatedly
until one has $n$ orthogonal vectors.  One then takes the logarithm of the
rate of change in each of those directions.  Allowing the frame to evolve
for many driving cycles lets one follow a phase space cell as it samples
all parts of an attractor.  In this way one calculates the average
values of the exponents.  The values of the $\lambda_i$ are global
quantities, characterizing the attractor as a whole, or equivalently, the
long-term behavior of orbits throughout the attractor.

Calculating the highest exponent, $\lambda_1$, is not very difficult.
Finding values for a full spectrum of exponents, however, is rather
tricky, and requires a subtle touch.  I refer those interested
to the papers of Wolf, et al. and Brown, et al. for details
\cite{Wolfetal,BrownBryAbar}.
These definitions for
fractal dimension and Lyapunov exponents run into trouble in the quantum
case, where taking limits as $\epsilon\rightarrow0$ is not very
well-defined.  We will see in section V how one can adjust these
definitions appropriately.


\section{{\bf QUANTUM MAPS AND STRANGE ATTRACTORS}}

\subsection{Quantum Maps.}

In the classical case we went from continuous to discrete dynamics by
going to the constant phase map of the Duffing oscillator.  In considering
a quantum equivalent, it is convenient to coarse-grain our selected
trajectories $x(t)$ and $\xp(t)$ by considering only their values at the
times $t_i = 2\pi i/\omega_0$ of constant phase.  The decoherence functional
then becomes
\begin{equation}
D[\{x_i\},\{\xp_i\}] = \int_{\{x_i\}} \delx \int_{\{\xp_i\}} \delxp
  D[x(t),\xp(t)],
\end{equation}
where the decoherence functional on continuous trajectories $D[x(t),\xp(t)]$
is given by (\ref{cont_decoherence}).  The path integrals are over all
paths of $x(t)$ and $\xp(t)$ which pass through the points $x_i$ and
$\xp_i$ respectively at times $t_i$.

Such a coarse-graining is discussed by Gell-Mann and Hartle \cite{GMHart3}.
In general,
in order for such a system to be sufficiently decoherent, we must also
coarse-grain on the positions $\{x_i\}$ and $\{\xp_i\}$.  Instead of
specifying the positions exactly, we instead require just that the positions
fall in one of a group of short intervals $\Delta^i_{\alpha_i}$ at the
times $t_i$.  A history is then given by specifying the sequence of
$\alpha_i$'s.  We'll use the short-hand notation $\alpha$ for this
sequence.

We can estimate the minimum length $d$ of such intervals by requiring
that the off-diagonal terms of the decoherence functional be strongly
supressed for $|x_i - \xp_i| = |\xi_i| > d$.  As Gell-Mann and Hartle
show, this depends on the separation between times $t_i$, the strength of
the coupling, and so forth.  A rough estimate gives
\begin{equation}
\Delta t \sim {1\over\Omega}\exp \biggl( {\pi\hbar\over{M\Gamma d^2}} \biggr).
\end{equation}
For the quantum maps the interval is $\Delta t = 2\pi/\omega_0$.  $\Omega$
is the frequency cutoff; it is equal to $K/M\Gamma$.  So we get an estimate
\begin{equation}
d^2 \sim {\pi\hbar\over{M\Gamma \ln(2 \pi K/M\Gamma\omega_0)}}.
\end{equation}

Since equation (\ref{cont_decoherence}) is expressed in terms of the
variables $X(t)$ and $\xi(t)$, it might be useful to change variables in
our coarse-grained systems.
As the decoherence functional is supressed for large $\xi$, we can
we can treat our integrand as being quadratic in $\xi(t)$, and
carry out the $\xi$ integration.  This gives us
\begin{equation}
p(\alpha) = \sqrt{2\pi\over K} \int_{\{\alpha_i\}} \delX \exp \biggl\{
  - {1\over{K\hbar}} \intof e^2(t) dt \biggr\} w(X_0,p_0),
\end{equation}
where $e(t) = 0$ is the classical equation of motion as given above in
(\ref{EOM}) and $w(X_0,p_0)$ is the initial Wigner distribution, obtained
by the integral over $\xi_0$.

We see that
if the $\{X_i\}$ do not lie along a classical trajectory $e(t) = 0$, then
the functional will be supressed.  So
the most probable histories are those which lie along the classical
trajectory.  The $X$ path integral cannot be done exactly in most cases,
but one can see that in general the $\{X_i\}$ must lie near the
$\{x_i\}$ for some classical problem for the probability to be of reasonable
magnitude.

The Wigner distribution $w(X,p)$ (not to be confused with the
influence phase $W[X(t),\xi(t)]$, which is a functional!) is given by
\begin{equation}
w(X,p) = {1\over\pi} \int_{-\infty}^{+\infty} e^{i\xi p/\hbar}
  \rhot(X+\xi/2,X-\xi/2) d\xi,
\label{Wigner}
\end{equation}
where we see that $X$ and $\xi$ are our usual variables, and $p$
has units of momentum.  $w(X,p)$ is somewhat analogous to a probability
distribution on phase space.  It is real, and integrates to a total of
1; its primary difference
from a classical phase space distribution is that
it is not strictly non-negative.

If we want to advance $w(X,p)$ in time, we can define a transfer matrix
${\bf T}$ such that $w(t_f) = {\bf T} w(t_0)$.  More explicitly
this is
\begin{equation}
w(X_f,p_f) = \int dX_0 \int dp_0\ T(X_f,p_f; X_0,p_0) w(X_0,p_0)
\end{equation}
where $T(X_f,p_f; X_0,p_0)$ is defined
\begin{mathletters}
\begin{equation}
T(X_f,p_f; X_0,p_0) = {1\over\pi} \int d\xi_0 d\xi_f
  \e^{i(\xi_f p_f - \xi_0 p_0)/\hbar}
  {\tilde T}(X_f + \xi_f/2, X_f - \xi_f/2; X_0 + \xi_0/2, X_0 - \xi_0/2),
\end{equation}
\begin{equation}
{\tilde T}(x_f,\xp_f; x_0, \xp_0) =
  \int \delx \delxp \exp {i\over\hbar}\biggl\{
  \Ss[x(t)] - \Ss[\xp(t)] + W[x(t),\xp(t)] \biggr\}
\end{equation}
\end{mathletters}
If we let $t_f - t_0 = 2\pi/\omega_0$ then time advancement can
be performed by repeated applications of ${\bf T}$.  This is a sort of
{\it quantum map},
\begin{equation}
w_i \rightarrow w_{i+1} = {\bf T} w_i.
\end{equation}
We can ask if repeated applications of ${\bf T}$ will tend to converge to
some invariant Wigner distribution $\winv = {\bf T} \winv$,
analogous to the invariant measure $\Pinv$ of section III.
Preliminary numerical calculations seem to show that
this is the case \cite{Brun2}.  This $\winv$ appears unique, and
should be analytic, thanks to the ``blurring'' effect of quantum noise.

To make closer contact with the classical system, we might wish instead
to consider histories in which a trajectory passes through small cells in
phase space, rather than just intervals in $X$.  We can write such a history
by considering projections onto intervals in $X$ followed very briefly by
projections onto intervals in $p$.  Histories of this type have been
considered by Gell-Mann and Hartle, and by Halliwell, who wrote down an
explicit equation for such a history \cite{GMHart3,Halliwell}.

One cannot in general specify both the momentum and position of a particle
at the same instant.  One can, however, consider a measurement of position
followed by a measurement of momentum, and let the time between them
go to zero.  Halliwell calculated the probability of such a history using
approximate projections
\begin{mathletters}
\begin{equation}
P_{\bar x} = {1\over{\pi^{1/2}\sigma_{\bar x}}} \int_{-\infty}^{\infty} dx
  \exp\biggl[ - {{(x - {\bar x})^2}\over{\sigma_{\bar x}^2}} \biggr]
  \ket{x}\bra{x},
\end{equation}
\begin{equation}
P_{\bar p} = {1\over{\pi^{1/2}\sigma_{\bar p}}} \int_{-\infty}^{\infty} dp
  \exp\biggl[ - {{(p - {\bar p})^2}\over{\sigma_{\bar p}^2}} \biggr]
  \ket{p}\bra{p}.
\end{equation}
\end{mathletters}
These are approximate projections into intervals $\Delta_{\bar x}$
of width $\sigma_{\bar x}$ and ${\tilde \Delta}_{\bar p}$ of width
$\sigma_{\bar p}$, centered on ${\bar x}$ and ${\bar p}$, respectively.
Halliwell shows \cite{Halliwell}
that for an initial Wigner distribution $w(X,p)$,
the probability of finding a particle in the phase-space cell delimited by
the two above projections is
\begin{equation}
p(\Delta_{\bar x},{\tilde \Delta}_{\bar p}) =
  \int dX dp\ w(X,p) \exp\bigl[ - a(p - {\bar p})^2 - b(X - {\bar x})^2 \bigr],
\label{Onetime}
\end{equation}
where
\begin{equation}
a = {{\sigma_{\bar x}^2}\over{2[\hbar^2
  + (1/4)\sigma_{\bar x}^2\sigma_{\bar p}^2]}},\
b = {2\over{\sigma_{\bar x}^2}}
\end{equation}
when the $X$ projection preceeds the $p$ projection, and
\begin{equation}
a = {2\over{\sigma_{\bar p}^2}},\
b = {{\sigma_{\bar p}^2}\over{2[\hbar^2
  + (1/4)\sigma_{\bar x}^2\sigma_{\bar p}^2]}}
\end{equation}
when the $p$ projection preceeds the $X$ projection.  There is a restriction
on these projections that
\begin{equation}
0 < ab \le {1\over\hbar} \Longrightarrow
  \sigma_{\bar x}^2\sigma_{\bar p}^2 > \hbar^2.
\end{equation}

To calculate the probabilities of an orbit passing through a series of such
cells $(\Delta_i,{\tilde\Delta}_i)$ at times $t_i$ we make use of the
transition
matrix ${\bf T}$.  Let us assume that the $X$ projection comes first.  Then
it turns out that
\begin{eqnarray}
p(\{ \Delta_i,{\tilde \Delta}_i \}) = &&
  \int d\{X\} d\{p\} w(X_0,p_0) \nonumber\\
&& \times \exp\bigl[ - {2\over{\sigma_{\bar p}^2}} (p_1 - {\bar p_0})^2
  -{2\over{\sigma_{\bar x}^2}} (X_0 - {\bar x_0})^2
  -{{\sigma_{\bar x}^2}\over{2\hbar^2}} (p_0 - p_1)^2
  -{{\sigma_{\bar p}^2}\over{2\hbar^2}} (X_0 - X_1)^2
  \bigr] \nonumber\\
&& \times T(X_2,p_2; X_1,p_1) \nonumber\\
&& \times \exp\bigl[ - {2\over{\sigma_{\bar p}^2}} (p_3 - {\bar p_1})^2
  -{2\over{\sigma_{\bar x}^2}} (X_2 - {\bar x_1})^2
  -{{\sigma_{\bar x}^2}\over{2\hbar^2}} (p_2 - p_3)^2
  -{{\sigma_{\bar p}^2}\over{2\hbar^2}} (X_2 - X_3)^2
  \bigr] \nonumber\\
&& \times T(X_4,p_4; X_3,p_3) \nonumber\\
&& \times \cdots
\label{Manytimes}
\end{eqnarray}
where the interval $\Delta_i$ is centered on ${\bar x}_i$ and
${\tilde\Delta}_i$ on ${\bar p}_i$.  The final projection will be of the
form (\ref{Onetime}).  The case where the $p$ projection precedes the
$x$ is very similar to the above.

There are other ways of considering phase space projections, using coherent
states or Gaussian combinations of coherent states.  A fuller discussion of
phase space histories and their decoherence deserves a fuller discussion
elsewhere \cite{Brun3}.

\subsection{The Master Equation.}

Another common method of studying systems such as this is by means of a
Master equation formalism.  Caldeira and Leggett derive such an equation
in the case of a harmonic oscillator interacting with a thermal bath at
relatively high temperature \cite{CaldLegg}.
Their result is readily adapted to the present
case, yielding the equation
\begin{eqnarray}
{{\partial\rhot}\over{\partial t}}(x,\xp) = &&
  - {K\over\hbar}(x - \xp)^2 \rhot
  + {iq\over\hbar}(x - \xp)\cos(\omega_0 t) \rhot
  - {i\over\hbar}(V(x) - V(\xp)) \rhot \nonumber\\
&& + 2\Gamma (x-\xp)\biggl( {{\partial\rhot}\over{\partial\xp}}
   - {{\partial\rhot}\over{\partial x}} \biggr)
   + {{i\hbar}\over{2m}}\biggl( {{\partial^2\rhot}\over{\partial x^2}}
   - {{\partial^2\rhot}\over{\partial\xp^2}} \biggr)
\end{eqnarray}
In examining this equation, the meanings of the different terms are highly
intuitive.  The $K/\hbar$ term is a diffusive effect resulting from the
quantum noise; the $\Gamma$ term includes the effects of dissipation; the
$q\cos(\omega_0 t)$ is the driving force.

Changing to the variables $X$ and $\xi$, the Master equation becomes
\begin{eqnarray}
{{\partial\rhot}\over{\partial t}}(X,\xi) = &&
  - (4 K/\hbar) \xi^2 \rhot +
  (2 i q/\hbar) \xi \cos(\omega_0 t) \rhot - {i\over\hbar} \bigl(
  V(X+\xi/2) - V(X-\xi/2) \bigr) \rhot \nonumber\\
&&  - 4\Gamma \xi {{\partial\rhot}\over{\partial\xi}}
  + {{i\hbar}\over{2m}}{{\partial^2\rhot}\over{\partial\xi\partial X}}.
\end{eqnarray}
If $\xi$ is small, then we can expand the potential term to give us
\begin{equation}
V(X+\xi/2)-V(X-\xi/2) \approx \xi{{\partial V}\over{\partial X}}(X)
  + {{\xi^3}\over24}{{\partial^3 V}\over{\partial X^3}} + \cdots.
\end{equation}
For the Duffing potential, of course, the higher-order terms vanish.
This is a convenient benefit of dealing with a polynomial potential.

Transforming this equation by (\ref{Wigner}) gives us a new equation
for the evolution of the Wigner distribution itself:
\begin{eqnarray}
{{\partial w}\over{\partial t}}(X,p) = &&
  \hbar K {{\partial^2 w}\over{\partial p^2}}
  - q\cos(\omega_0 t){{\partial w}\over{\partial p}}
  + {{\partial V}\over{\partial X}}(X){{\partial w}\over{\partial
p}}\nonumber\\
&&  - {{\hbar^2}\over{24}}{{\partial^3 V}\over{\partial X^3}}(X)
    {{\partial^3 w}\over{\partial p^3}}
  + 4\Gamma{\partial\over{\partial p}} p w
  - {p\over m}{{\partial w}\over{\partial X}}.
\end{eqnarray}
This is almost exactly the form of the
Fokker-Planck equation for the classical equation of motion (\ref{EOM}),
with the diffusive $\hbar K$ term representing the effects of the
random fluctuations on the ``probability'' distribution  and the
third-derivative term being a purely quantum-mechanical addition,
enabling $w(X,p)$ to become negative in limited regions of phase space.
To interpret this distribution as a probability, we must
coarse-grain by averaging it over small volumes of phase space, producing
the sort of ``smeared'' Wigner distribution discussed by Halliwell
\cite{Halliwell}.

This is not, of course, the full story.  In order to correctly describe
this system, we need not only the time evolution of the Wigner distribution,
but also to specify a set of decoherent histories, as discussed in the
previous section.  Without those histories, it is impossible to assign
classical probabilities in a consistent manner.
These two approaches can be made
to complement each other, however, as the Master equation can be solved
to yield the transfer matrix ${\bf T}$, defined in the previous section as
a path integral.  In the case of chaos, one can in general only solve
these equations numerically, and the Master equation formalism
then has a compuational advantage over the path integral form.

\section{{\bf INTERPRETATION OF CLASSICAL QUANTITIES}}

{}From the previous section, we see that the behavior of a system such as we
are examining can be evaluated on many levels:

1.  The Classical level.  In the previous section we saw that all histories
which deviate too far from the classical solution have their probabilities
highly suppressed.  If a system is large enough in scale, with enough
inertia that the quantum effects are lost in other sources of uncertainty,
we can treat it as approximately classical.  Clearly, in a chaotic system
this quantum noise does cause large alterations in the overall
behavior of the system, but it is often impossible to separate this from
thermal noise or other sources of error.

2.  The Quasiclassical level.  Here we again treat the system as essentially
classical, but now explicitly include the noise arising from quantum effects,
which is large enough to be noticed on the scale of the system; this is
the system as described by equation (\ref{EOM}).  From a practical point
of view, this is the level at which quantum effects are most easily
calculated.  This also overlaps the considerable work that has
been done on dynamical systems with
noise \cite{Wolfetal,ZipLuc,OttHanson,RechWhite}.  Note that this is {\it not}
the same as a semiclassical approximation, such as WKB.

3.  The Quantum level.  Fundamentally, we can consider the system in terms
of coarse-grainings and decoherent histories.  Instead of treating a system
as basically classical with added noise, we consider all possible histories,
and compute expectation values for classical quantities from the
probabilities of those histories.

We've already discussed the classical (level 1)
definitions of the lyapunov exponents and fractal dimensions
used to characterize chaotic systems and strange attractors in section III.
As pointed out,
these quantities are usually defined at least formally by
calculating a quantity for the system at different levels
of coarse-graining (i.e., different box sizes $\epsilon$), and taking the limit
as we go to finer and finer scales.
While this has great mathematical power and consistency, in actual
physical systems it inevitably breaks down.
As Benoit Mandelbrot wrote
on the problem of measuring coastlines with seemingly infinite levels of
detail \cite{Mandelbrot},

``To obtain a [fractal] Koch curve,
the cascade of smaller and smaller new promontories
is pushed to infinity, but in Nature every cascade must stop or change
character.  While endless promontories may exist, the notion that they are
self-similar can apply only between certain limits.  Below the lower limit,
the concept of coastline ceases to belong to geography.

``It is therefore reasonable to view the real coastline as involving two
{\it cutoff scales}.  Its {\it outer cutoff} $\Omega$ might be the diameter
of the smallest circle encompassing an island, or perhaps a continent, and
the {\it inner cutoff} $\epsilon$ might be the twenty meters mentioned\ldots
Actual numerical values are hard to pinpoint, but the need for
cutoffs is unquestionable.''

As we shall see, in the case of chaotic strange attractors, the underlying
quantum physics effectively provides that lower cutoff.

\subsection{Lyapunov Exponents.}

Classically, the Lyapunov exponents characterize the rate at which nearby
trajectories diverge as they evolve according to the equations of motion.
In a chaotic system, one expects any two trajectories, no matter how close
they start, to eventually move on the strange attractor completely
independently of each other.  This is measured in the classical case by
taking the limit as points start arbitrarily near each other and evolve
for arbitrarily long lengths of time.

When we allow for the presence of quantum effects, however, this definition
is no longer meaningful.  As points begin closer and closer to each other,
the effects of noise become larger and larger; one would expect the largest
exponent to diverge in the limit as $\epsilon\rightarrow0$.  As we saw in
section IV, the phase space cells in a decoherent history cannot be smaller
than a certain size.  This limit provides the lower cutoff mentioned above.

For most systems it is impossible to calculate the values of Lyapunov
exponents exactly.  Instead, one performs a numerical calculation.
It is easiest to calculate the highest exponent; lower exponents are more
difficult, as their effects tend to be swamped by $\lambda_1$.  In a numerical
calculation small errors are unavoidable; each such error will add a small
admixture of the most rapidly growing component, which will quickly
drown out other effects.

Because of this, we'll first consider only the value of $\lambda_1$.  A
simple way of estimating $\lambda_1$ is to numerically integrate
equation (\ref{EOM}) for a longer period of time, to generate a large
number of points $\{x_i,p_i\}$ in phase space.  One can then locate
nearby points, closer than a certain cutoff $\epsilon$, and trace their
trajectories until they diverge further than an upper cutoff $\Delta$.
One then calculates the logarithm of the average divergence rate and
averages it over many such pairs of points.

I have calculated this quantity in the quasiclassical case (see fig.~4).
It turns out that the result one calculates is not very sensitive to
the upper cutoff $\Delta$, but is highly sensitive to the lower cutoff
$\epsilon$.  In figure 4 we see the measured value of $\lambda_1$ as
a function of $\epsilon$ for several different relative strengths of the
quantum noise $\hbar K$.  Notice how for $\hbar K > 0$ $\lambda_1$
diverges as $\epsilon \rightarrow 0$.

Because of dissipation, the overall phase-space volume of an initial
distribution tends to decrease with time.
This indicates that, classically,
$\lambda_1 + \lambda_2 < 0$.  At very small length
scales, however, the effects of noise counteract the effects of dissipation,
causing phase-space volume to grow rather than shrink.  Thus, at small
length scales we expect to see the sum $\lambda_1 + \lambda_2$ become
positive, and eventually approach $\lambda_1/\lambda_2 \approx 1$; the
dimension
at that length scales should also approach an integer (2 in this case).

We can try to define a quantum-mechanical analog of $\lambda_1$.  While I
am not sure exactly what form such a definition should take, I can make
a conjecture.  Suppose that we start from the invariant Wigner distribution
$\winv$ at $t=t_0$.
We divide phase space into small cells $\{c_i\}$, centered on average
positions $\{{\vec v}_i\}$ in phase space.  These cells have a characteristic
size $\epsilon$ (or area $\epsilon^2$).  Let $d_{ij}$ be the distance between
the centers of the $i$th and $j$th cells.  We can define $p_i$ to be the
probability that the system is in cell $c_i$ at time $t_1$, using equation
(\ref{Onetime}), and $p_{ij}$ to be the probability that the system is in
$c_i$ at time $t_1$ and $c_j$ at time $t_2$, as shown in (\ref{Manytimes}).
Clearly
\[
\sum_i p_i = 1,\ \ \ \sum_j p_{ij} = p_i.
\]
The probability of the system being in $c_j$ at time $t_2$
{\it given} that it was in
$c_i$ at $t_1$ is
\begin{equation}
p(j|i) = {{p_{ij}}\over{p_i}}.
\end{equation}
A rough estimate of the rate of spreading is then given by
\begin{equation}
\lambda(\epsilon)_{\rm qm} = {1\over2} \sum_i p_i \log \left[ {
  {{1\over2}\sum\limits_{j,k} p_{ij} p_{ik} d_{jk}^2}\over{p_i^2\epsilon^2}
  } \right].
\end{equation}

It isn't clear whether this will agree with the usual definition of
$\lambda_1$ in the limit.  This is much more a rate of expansion averaged
over the attractor, whereas $\lambda_1$ is usually defined as the rate
at which nearby solutions diverge when followed for a long period of time.
This latter definition has serious problems in the quantum case, where it
is impossible to start solutions arbitrarily close together, and hence
equally impossible to follow them for arbitrarily long periods of time
without global processes (such as folding) becoming important.  When quantum
effects are very small one can approach this long-orbit definition,
but in that case one is really doing
a quasiclassical calculation (like the one above in figure 4),
where the system can be
treated as a classical stochastic equation.

Numerical experiments might serve to explore the connections, if any,
between these classical and quantum ideas of Lyapunov exponents.  I hope
to do more such exploration soon.  Also, it is not clear to me exactly
what form quantum equivalents to lower Lyapunov exponents might take, nor
even if such a concept is useful.  These questions will soon, I hope, have
at least tentative answers.

\subsection{Information Dimension.}

The information dimension is, as we saw in section IV, another number used
to characterize strange attractors.  It is usually defined by a box-counting
algorithm of the type given in (\ref{infodim}).  As mentioned before, when
one takes quantum mechanics into account, allowing the size of a box to
go to zero no longer makes much sense.

Instead, let us consider the information dimension $D_I$ as a function of
box-size:
\begin{equation}
D_I(\epsilon) = {{\sum\limits_i p_i\ln p_i}\over{\ln\epsilon}}.
\label{D_of_eps}
\end{equation}
The probability $p_i$ is defined as before.  For non-zero $\epsilon$,
the exact value of $D_I(\epsilon)$ will vary slightly depending on how
the boxes are chosen.  This ambiguity can be eliminated by taking
$D_I(\epsilon)$
to be the minimum possible value over all possible arrangements of boxes.
In practice, this makes little difference.  The usual classical limit
is then just the limit of $D_I(\epsilon)$ as $\epsilon \rightarrow 0$.

Figure 5 shows the calculated values of $D_I(\epsilon)$ for different
values of $\hbar K$.  As we see, at large length scales the fractal
nature of the attractor is not readily apparent; as we shrink our
scale, the dimension decreases, until when dropping below the lower
cutoff given by the quantum effects it abruptly turns upward again.  This
was calculated quasiclassically, using a long orbit with associated noise.

One can also do an analogous calculation using the complete quantum
theory.  Consider the invariant Wigner distribution $\winv(X,p)$, defined
in section IV.  We can define the information dimension $D_I(\epsilon)_{\rm
qm}$
using the same definition (\ref{D_of_eps}).
We divide phase space into evenly-sized cells $\{c_i\}$
of size $\epsilon$, just as in the discussion of
$\lambda(\epsilon)_{\rm qm}$ above,
and use the expression (\ref{Onetime}) for the probability $p_i$ of being in
the
cell $c_i$.  Again, we can eliminate ambiguity by minimizing
$D_I(\epsilon)_{\rm qm}$ over all possible divisions into cells.  Clearly such
a
dimension will not even be well defined for cells of volume less than
$\hbar$, and will in general depend on the scale of the coarse-graining, just
as in the quasiclassical treatment.

Needless to say, it is much easier to extend the information dimension $D_I$
to a probabilistic theory than it is to find an analogy for the capacity
dimension $D_C$.  The presence of noise will give a small but non-zero
probability of finding a point in any cell, even if it is far from the
classical
strange attractor.
So, for quantum dissipative chaos at any rate, $D_I$ seems
to be the more useful quantity.

\section{{\bf QUANTUM CHAOS}}

Since the discovery of chaos in the 1970's, there have been numerous
attempts to look for the existence of chaos in quantum mechanical systems.
Almost all of these have concentrated on quantized versions of non-integrable
Hamiltonian systems.  A seeming paradox was at the heart of the debate:
chaos, as a classical phenomenon, depended entirely on the existence of
nonlinear terms in the equations of motion; yet quantum systems are by their
nature completely linear.

In fact, this argument is clearly invalid.  While it is true that
the linearity of the Schr\"odinger equation and its relativistic
generalizations implies that one would not expect chaotic behavior in
the wave function itself, this has little bearing on what one would
actually see if one observed such a system.  One does not measure wave
functions; one measures particles.

An analogous classical treatment would be to consider probability
distributions in phase space rather than values of position and momentum.
One can then go from a set of nonlinear ordinary differential equations
to a Fokker-Planck partial differential equation.  The P.D.E. is completely
linear.  Does this then imply that chaos cannot exist in classical mechanics?
Such a conclusion would be absurd.

Of course, wave functions are not probability distributions, so the comparison
is a bit misleading.  A sufficiently coarse-grained Wigner distribution,
however, can be made to look very much like a probability distribution, and
its Master equation closely resembles the Fokker-Planck equation, as we have
seen, so comparing the two is not completely inappropriate.

In fact, certain quantum models can
exhibit chaos \cite{Chiretal}, but they are exceptional.  Most systems
which occupy a bounded volume in phase space do eventually exhibit
{\it quantum recurrence} in which the expectation values of quantities
such as energy are almost periodic
\cite{FishGremPrange,HoggHub,CasChirShep,FordIlg}.  A simple information-theory
argument can be made for why this should be:  a bounded volume in phase
space $V$ represents a finite number of possible states $N \sim V/\hbar$,
so one would expect the long-term behavior to be periodic or quasiperiodic.
According to Ehrenfest's theorem, a narrow wave packet will tend to follow
the classical chaotic trajectory for a time $t_E$ until it has spread out
to a size comparable to the phase space volume.  Thus, it is argued, one
should observe a long chaotic ``transient,'' ending ultimately in periodic
or almost periodic motion.  As the system becomes more ``macroscopic''
the phase space volume becomes larger with respect to $\hbar$ and
$t_E$ becomes longer.

Chirikov, Izrailev and Shepelyansky describe the usual approach to quantum
chaos \cite{Chiretal}.  They separate the problem into two parts, the
dynamics of the undisturbed wave function and the effects of measurement
and wave function collapse.  Obviously, this periodic behavior of expectation
values says nothing about what an experimenter would actually observe upon
measuring the system.  They compare the former with deterministic behavior
and the latter with randomness and ``noise''.  Clearly, an actual series
of measurements would not be periodic at all, but would instead resemble
a random chaotic trajectory.  They further dismiss the study of dissipative
chaos as a mere phenomenological approximation to an underlying Hamiltonian
system, e.g., in our case including both the system and reservoir degrees of
freedom.

If the work of Gell-Mann, Hartle, and others is to be believed, we
should consider only decoherent histories.  Since Hamiltonian systems
almost by definition do not interact with outside degrees of freedom,
one cannot really talk of ``measuring'' them.  If measurements are taken,
the system is disturbed; if measurements are not taken, the system evolves
undisturbed, but detailed histories of the motion will not decohere.  In
dissipative systems, by contrast, the chaotic system is interacting continually
with the neglected degrees of freedom of the reservoir.  These serve to
provide a continual ``measurement'' of the system, in addition to causing
dissipation and noise, so that histories of the system variables do decohere,
as we have seen.  Thus, considering these dissipative systems from a
decoherence functional point of view is entirely appropriate.  Also, any
description of real macroscopic systems must allow for coarse-graining
over the many degrees of freedom of which we are ignorant.  As one goes
from the classical to the quantum realms, deterministic uncertainty or
randomness is replaced by probabilistic uncertainty.

In the Duffing oscillator model studied here, an
arbitary initial distribution $w_0$ will tend to converge onto the invariant
distribution, $\lim\limits_{n\rightarrow\infty}{\bf T}^n w_0 \rightarrow \winv$
\cite{Brun2}.
This invariant distribution is periodic, with period $2\pi/\omega_0$,
so that ${\bf T}\winv = \winv$.  This resembles the idea of a long
chaotic ``transient'' leading to a non-chaotic, periodic behavior.

This is not, however, very different from the behavior of classical systems
evolving according to the Fokker-Planck equation.  Most initial distributions
$P_0$ of non-zero width evolve into the invariant probability measure $\Pinv$
in much the same way that $w$ evolves into $\winv$ in the quantum case
(see fig. 6).  The
classical case is complicated by the non-analyticity and non-uniqueness of
$\Pinv$, so that there can be a finite probability of sitting on top of some
unstable equilibrium or periodic point.  An initial distribution with non-zero
width will never actually become the invariant measure, of course, always
itself
remaining an analytic function if it begins as one, but the difference
rapidly becomes too small to measure.  These difficulties do not exist in
the quantum case, due to the presence of noise.  This difference, plus the
necessity for coarse-graining (which requires an initial distribution to
have non-zero width, thanks to the uncertainty principle),
is what distinguishes the classical and quantum
cases, not the periodicity of the solution per se.  The important idea is
not the behavior of $\winv$, but rather the probabilities of different
possible histories of the system as described in (\ref{Manytimes}).

Other authors have concentrated on the rate of expansion of uncertainty in
quantum chaotic systems, and on how chaos leads to a form of ``dissipation''
in the quantum wave function itself \cite{FoxLan,Fox,Boncietal}.

A very interesting
suggestion arose in work by Weinberg on possible nonlinear generalizations
of quantum mechanics \cite{Weinberg1,Weinberg2}.  With true nonlinearity,
chaos on the level of the wave function could exist.  Weinberg points out,
however, that any nonlinearity in the theory must be very small to be
consistent with experiment, and chaos is rare in such nearly-integrable
systems.  While such a generalization of quantum theory is certainly not
ruled out be experiment, it poses a number of troubling problems.  As
pointed out by Polchinski \cite{Polchinski}, either
faster-than-light transmission of
information via an EPR-type experiment becomes possible, or different
branches of histories can continue to interact with each other, effectively
making decoherence impossible.  Both of these, while
certainly not impossible, have absolutely no basis in experimental evidence;
indeed, the latter would imply that even classical probabilities could
undergo a sort of ``interference'' with each other.

A few papers have been published on quantum dissipative
chaos \cite{DavidSanth,NorenMilek},
mostly by including a phenomenological term in the Schr\"odinger equation
to model dissipation.  While there is a fair bit of other work on the
phenomenon of quantum dissipation itself
\cite{CaldLegg,FeynVern,Davidson,Vitiello}, little of it has
been applied to chaotic systems, as far as I am aware, and relatively little
has been from the decoherence point of view.

I cannot do a broad survey of an extremely active field here.  A
number of good books now exist on quantum Hamiltonian chaos
\cite{Gutzwiller,GianVorosZinn}.

\section{{\bf CONCLUSIONS}}

In this paper, I have only treated a single model in depth:  the forced,
damped Duffing oscillator.  It is clear, though, that the techniques used
are easily applied to any nonlinear oscillator problem.  I believe that,
in general, the formalism of Gell-Mann and Hartle provides a rigorous
method for treating any classical chaotic system.

For some systems, of course, such a treatment may be inappropriate.
Dissipative
chaos was first discovered in attempts to model fluid dynamics; while in
principle such problems could be treated in this way, any quantum effects
are likely to be so small as to be irrelevant.  For nonlinear oscillators,
though, the application is quite reasonable, and it is possible that
experiments could be done in quantum optical or electronic systems which
would correspond to classically chaotic systems of this sort.

What is more, using this formalism provides a link between a
quantum system and its ``classical limit'' which lets one define the idea
of quantum chaos in a rigorous way.  And such systems can be treated not
only semi-classically, but with the full quantum laws as well.

Equally important, in this quasiclassical treatment we can see how chaotic
systems can serve to amplify the effects of small quantum fluctuations.
Sensitive dependence on initial conditions -- the hallmark of chaos --
is an important idea in understanding {\it measurement situations}, in which
minor quantum effects can become correlated with a change in macroscopic
variables.  In complex systems with many degrees of freedom, sensitive
dependence on initial conditions is probably the rule rather than the
exception.

A great deal remains to be done.  This formalism can readily be applied to
a number of other systems besides the Duffing oscillator.  Also, it would
be valuable to develop numerical programs for solving the Master equation,
and calculating the quantities characterizing the attractor in the full
quantum system as well as the quasiclassical limit.  Some work on this has
already been done \cite{Brun2}, but it is still in a rather crude state.
But the basic outlines of the theory
are clear.  Using the theory of Gell-Mann and Hartle, a rigorous treatment
of quantum dissipative chaos is finally possible.

\acknowledgements

I gratefully acknowledge the invaluable support of Murray Gell-Mann and
Jim Hartle, without whom this paper could not have been written.  Thanks
to Seth Lloyd for many stimulating and helpful discussions, to
Michael Cross and Sima Setayeshgar for assistance with aspects of the
classical chaos theory and pointing out a number of the references, and
to Reggie Brown for providing me with assistance in numerical calculations.
Finally, thanks to Jonathon Halliwell, for his interest in my work and
his encouragement.  This work was supported in part by DOE Grant
No. DE-FG03-92-ER40701.


\begin{references}
\bibitem{GMHart1} M. Gell-Mann and J. Hartle, in {\it Complexity,
  Entropy, and the Physics of Information}, SFI Studies in the
  Sciences of Complexity v.8, ed. by W. Zurek, Addison-Wesley,
  Reading (1990).
\bibitem{GMHart2} M. Gell-Mann and J. Hartle, Proceedings of the 25th
  International Conference on High-Energy Physics, Singapore, 1303
  (August 1990).
\bibitem{GMHart3} M. Gell-Mann and J. Hartle, Phys. Rev. D {\bf 47},
  3345 (1993).
\bibitem{Brun} T. Brun, ``Quasiclassical equations of motion for nonlinear
  Brownian systems,'' to appear in Phys. Rev. D 1993.
\bibitem{Merzbacher} E. Merzbacher, {\it Quantum Mechanics}, 2nd Edition,
  John Wiley \& Sons, Inc., New York (1970).
\bibitem{Zwanzig}  R. Zwanzig, J. Stat. Phys., {\bf 9}, 215 (1973).
\bibitem{CaldLegg} A.O. Caldeira and A.J. Leggett, Physica {\bf 121A},
  587 (1983).
\bibitem{GuckHolmes} J. Guckenheimer and P. Holmes, {\it Nonlinear
  Oscillations, Dynamical Systems, and Bifurcations of Vector Fields},
  Springer-Verlag New York Inc. (1983).
\bibitem{Wolfetal} A. Wolf, J. Swift, H. Swinney, J. Vastano, Physica
  {\bf 16}D, 285 (1985).
\bibitem{BrownBryAbar} R. Brown, P. Bryant, H. Abarbanel, Phys. Rev. A,
  {\bf 43}, 2787 (1991).
\bibitem{Brun2} T. Brun, unpublished.
\bibitem{Halliwell} J. J. Halliwell, Phys. Rev. D, {\bf 46}, 1610 (1992).
\bibitem{Brun3} T. Brun, unpublished.
\bibitem{ZipLuc} A. Zippelius and M. L\"ucke, J. Stat. Phys., {\bf 24},
  345 (1981).
\bibitem{OttHanson} E. Ott and J. Hanson, Phys. Lett., {\bf 85}A, 20
  (1981).
\bibitem{RechWhite} A. B. Rechester and R. B. White, Phys. Rev. A,
  {\bf 27}, 1203 (1983).
\bibitem{Mandelbrot} B. Mandelbrot, {\it The Fractal Geometry of Nature},
  W. H. Freeman and Company, New York (1977).
\bibitem{Chiretal} B. V. Chirikov, F. M. Izrailev, D. L. Shepelyansky,
  Physica D, {\bf 33}, 77 (1988).
\bibitem{FishGremPrange} S. Fishman, D.R. Grempel, R.E. Prange,
  Phys. Rev. Lett., {\bf 49}, 509 (1982).
\bibitem{HoggHub} T. Hogg and B. A. Huberman, Phys. Rev. Lett.,
  {\bf 48}, 711 (1982).
\bibitem{CasChirShep} G. Casati, B.V. Chirikov, D.L. Shepelyansky,
  Phys. Rev. Lett., {\bf 53}, 2526 (1984).
\bibitem{FordIlg} J. Ford and M. Ilg, Phys. Rev. A, {\bf 45}, 6165 (1992).
\bibitem{FoxLan} R.F. Fox and B.L. Lan, Phys. Rev. A, {\bf 41}, 2952 (1990).
\bibitem{Fox} R.F. Fox, Phys. Rev. A, {\bf 41}, 2969 (1990).
\bibitem{Boncietal}  L. Bonci, R. Roncaglia, B. West, P. Grigolini,
  Phys. Rev. A, {\bf 45}, 8940 (1992).
\bibitem{Weinberg1} S. Weinberg, Phys. Rev. Lett., {\bf 62}, 485 (1989).
\bibitem{Weinberg2} S. Weinberg, Ann. Phys., {\bf 194}, 336 (1989).
\bibitem{Polchinski} J. Polchinski, Phys. Rev. Lett., {\bf 66}, 397 (1991).
\bibitem{DavidSanth} A. Davidson and P. Santhanam, Phys. Rev. B,
  {\bf 46}, 3368 (1992).
\bibitem{NorenMilek} W. N\"orenberg and B. Milek, Nucl. Phys.,
  A{\bf 545}, 485 (1992).
\bibitem{FeynVern} R.P. Feynman and F.L. Vernon, Jr., Ann. Phys. {\bf 24},
  118 (1963).
\bibitem{Davidson} A. Davidson, Phys. Rev. A, {\bf 41}, 3395 (1990).
\bibitem{Vitiello} E. Celeghini, M. Rasetti, G. Vitiello, Ann. Phys.
  {\bf 215}, 156 (1992).
\bibitem{Gutzwiller} M.  Gutzwiller, {\it Chaos in Classical and Quantum
  Mechanics}, Springer-Verlag New York Inc. (1990).
\bibitem{GianVorosZinn} M.-J. Giannoni, A. Voros, J. Zinn-Justin, ed.,
  {\it Chaos and quantum physics},
  Elsevier Science Publishers B.V. (1991).
\end{references}
\end{document}